# A methodology for internal Web ethics

Michalis Vafopoulos
National Technical University of Athens
9 Heroon Polytechneiou st.
15773, Zografou Campus, Greece
+302107722538
vaf@aegean.gr

Petros Stefaneas
National Technical University of Athens
9 Heroon Polytechneiou st.
15773, Zografou Campus, Greece
+302107721869
petros@math.ntua.gr

Ioannis Anagnostopoulos
University of Central Greece
2-4 Papasiopoulou st.
35100, Galaneika – Lamia, Greece
+302231066937
janag@ucg.gr

Kieron O'Hara
University of Southampton
University campus, Southampton
SO17 1BJ, United Kingdom
+4402380592582
kmo@ecs.soton.ac.uk

## ABSTRACT
The vigorous impact of the Web in time and space arises from the fact that it motivates massive creation, editing and distribution of information by Users with little knowledge. This unprecedented continuum provides novel opportunities for innovation but also puts under jeopardy its survival as a stable construct that nurtures a complex system of connections. We examine the Web as an ethics determined space by demonstrating Hayek's theory of freedom in a three-leveled Web: technological, contextualized and economic. Our approach accounts for the co-dependence of code and values, and assumes that the Web is a self-contained system that exists in and by itself. This view of internal Web ethics directly connects the concept of freedom with issues like centralization of traffic and data control, rights on visiting log file, custom User profiles and the interplay among function, structure and morality of the Web. It is also demonstrated, in the case of Net Neutrality, that generic freedom-coercion trade-offs are incomplete in treating specific cases at work.

## Categories and Subject Descriptors
WSSC: "webscience.org/2010/E.4.3 Ethics in the Web"

## 1. INTRODUCTION
The amount of information on the Web is growing exponentially. Only in YouTube, 48 hours of video are uploaded every minute or nearly 8 years of content per day. Users' demands for a fast, secure, reliable, all-inclusive, trustworthy and general-purpose Web are uncontrollable. In 2010, the top 10 Web sites accounted for about 75 percent of all US traffic, compared to the 31% in 2001. Business controversies on issues like the monetization of links and excessive market power in searching and mobile applications are coming to the fore, whilst novel business models are changing the market rules (e.g. peer production, crowdfunding). Some executives and interest groups are still trying to conquer the Web by limiting the freedom to connect and update its content. Controversies have been also transferred to the legal battlefields. Contentious legal initiatives (e.g. SOPA) are causing both small and gigantic power games among governments, industries and non-governmental organizations. Concerns about identity, privacy and security are more often in the headlines. Although technically right solutions exist, these are have not been adopted yet (e.g. PKI, P3P, eID). HTML5 seems to gain interest well beyond technological outsets, and Open Data initiatives are revolutionizing science, business and government. Diverse debates and discussions are indirectly or directly connected to the Web ecosystem and outspread across the social discourse. Symbolically, all these issues are gathered under the rhetoric of online access as an emerging universal human right. Lately, national constitutions have started to incorporate it as a basic right (e.g. Norway). Internet and Web pioneers share different views on the issue, thus driving a creative dialogue about our live with the Web. This dialogue has raised, in various different ways and on as many different occasions, the following question: what kind of Web is more beneficial for society? Surely, as the transformational impact of the Web across society grows, the pressure to define its technological principles and the underlying moral values will escalate. Otherwise, we run the risk of ending up with a restrained, fragmented and autistic Web.

## 2. THE NEED FOR WEB ETHICS
Generic questions about Web's transformational potential have been brought into the agenda of many disciplines. Philosophical thinking and engineering should be in the front line by forming the main questions and setting the research framework. On this campaign single-sided analysis (i.e. technological or social) is not sufficient to tackle these complex and multifarious issues. Domain-specific analysis should be orchestrated by more generic approaches, expanding the solution range. Having defined existence, time and space in the Web [36], the next relevant quest is to consider its moral aspects.

Ethics is the branch of philosophy that deals with the study of good and evil. Its fundamental questions are often repeated through time, adaptive to the historical and social conditions. These fundamental questions include the definition of good and evil, the relation between morality and truth, the limits of freedom of will, the definition of right and wrong etc. Applied ethics is the branch of philosophy concerning the application of ethics to specific problems or classes of problems. From 1960's till today the field of applied ethics has seen remarkable growth. Business ethics, biomedical ethics, computer ethics, animal rights and environmental ethics are some of the most active areas in modern applied ethics [21]. The Vietnam War, the great progress in technology, the wide spread of drugs and contraceptives, the degradation of the environment, have raised a series of questions that could not be answered by traditional theories of ethics. An important contribution of applied ethics to the field of law is "A Theory of Justice" [33]. Computer ethics is the branch of applied ethics, which examines the social and ethical impact of information technology [18, 23]. More particularly, it focuses on the social impact that information technology has on our lives, the nature of such impact, and the utilization of technology in an ethical manner. Examples include issues related to the cybercrime, the protection of privacy, copyright and patents, the digital divide, and the use of computers in the workplace. The variety of technological applications creates new and unexpected situations and possibilities, thus causing new ethical dilemmas and values to emerge. For example, protection of personal data by electronic devices is of particular relevance to our society - to remember that only fifteen years ago the relative sensitivity was rudimentary. Lately, information ethics [12, 13] shed new light on many traditional ethical issues in computing.

The last twenty years there is a growing literature on the study of the ethics of the cyberspace encompassing all kinds of interactions among Users and the Internet [14, 31, 35]. Indicative topics include the ethics of blogging, free Speech and anonymity, pornography, censorship, intellectual property, privacy and regulation, spam and advertising, Internet as a medium of communication, accountability and trust, hacking, and the Internet access as a basic human right.

The Web has been built on the Internet stack, enabling the inter-linkage of digital beings. Despite the fact that it shares some common characteristics with its underlying technologies, creates a new feasibility and actuality space. The Web is sufficiently unusual, transformative and necessary to human existence, and as such it requires more systematic philosophical thinking to describe its ethically-relevant properties [28]. Initial motivation behind the development of the Web was based on ethical principles like esteem, pride, excellence, absence of guilt, rewards, and indignation [28]. Originally it was more a closed "Aristotelian world" than a space governed by rules, roles, hierarchies and deliverables. We believe that the above-mentioned virtues are the core driving forces of its exponential impact. These classic values that inspired the inventor and early Web Users and supported its massive dissemination, have now become more specific in practice. For instance, the discussion about freedom of expression incorporates the issue of Net Neutrality and self-determination that is connected to the privacy of online data.

One of the first questions for Web ethics should be a more comprehensive identification of the values that motivated its creation. An open conjecture in this line of inquiry has to do with the question whether different magmas of values and code could initiate similar decentralized information systems. Another question is how these evolving values affect the impact of Web in diverse social contexts and under what sort of prerequisites they can be sustainable.

It is now the time for scholars to look deep in the heart of the Web creation, to propose and engineer perspicacious solutions that will benefit the entire society. The quest for new requirements should directly address the needs, and promote human values. Web ethics should be thoroughly investigated in order to become a handy compass for Users, entrepreneurs and governments to direct their decisions towards prosperous ways.

## 3. INTERNAL WEB ETHICS

Web has been evolved from a piece of software code to a dynamical ecosystem of Users and multi-purpose functionalities. Despite its profound importance, Web ethics is still an unexplored research field. As such, it requires systematic research by determined experts.

The core of our methodology consists of two parts, firstly, the historical evolution of the Web and, secondly, a three-leveled approach thereof as this will be introduced below. The Web in its early stages was meant to address mainly technological needs, such as an interlinked bulleting board with low levels of interaction. In subsequent years, though, the Web evolved and became a construct of multiple interlocking contexts, and was even used to enable financial transactions. Users not only post and link digital content, but also communicate, comment, work, advertise, exchange information and physical goods in and through the Web. The social aspects of the Web are fashioned as the ability to create contexts, and an important part of them, economic contexts. Intense social and economic online transactions result into a dynamic magma of values and code. This fundamental standard implies that Web ethics should be studied under the assumption of inherent codependence between User and System (or equivalently Actor and Network [22]). Also, a sound definition regarding existence, time and space is necessary to describe the moral values tied to the Web as a system [36] In order to focus on our the methodology we propose in this paper, we assume that the Web is the only system existing in the universe ("manna from heaven" hypothesis"). Let us call this methodology the *internal Web ethics* analysis. Our approach extents the Web science perspective, which investigates the Web as a self-standing techno-social artifact [5, 38].

### 3.1 Magma of Users and code
Till the mass dissemination of Web 2.0, the main point in the ethics of computation took for granted that there was clear distinction between the technological and the social methodologies analyzing related phenomena. Technology was considered an autonomous force that changed society, and its methodology had a simple cause/effect form (technological determinism). Others believe the opposite, i.e., society is an autonomous force that changes technology (social determinism). Web 2.0 created a de facto indissoluble magma of Code and Users (techno-social evolution). Hence, the classic technology-society division is irrelevant in capturing the essence of the active User participation. The evolving interdependence between Code and Users can be addressed by models, which are built on the co-dependence of human moral values and engineering principles.

### 3.2 Being, time and space in the Web
Applied ethics methodologies refer to well-defined application domains. We believe that for the purposes of Web ethics a suitable framework is the definition of Web space [36]. A theory about existence in time and space is necessary to frame a tractable approach for the moral analysis of the Web. In [36] it has been proposed a notion of existence in the Web, based on a pragmatic definition of Being in general: *"a Being exists if and only if there is a communication channel linking to it"*. Being in the Web implies that the communication channel is concrete, identifiable and visible. Uniform Resource Identifier (URI) is the most profound and stable technology about creating communication channels in the Web. It requires the minimal description of invariant elements in communication through the Web and acts like the "fingerprint" of the Web Being because it is directly connected to existence (birth, access, navigate, edit and death). Thus, a Web Being is defined as follows: *"Web beings are defined to be Beings that can be communicated through the Web."*. The source of value for Web Beings is concentrated on how *digitality* is mutated by the *linking* potential, enabling them to be anywhere, at anytime. Users are "potential" owners of every Web Being, in the sense that the Being may not reside in the hardware but can be downloaded almost instantly. This expansion of the concept of existence is captured by the concept of *virtualization*, which describes the augmented potentialities of Web Being as a digital and distributable unity. The Web Space could be considered as a division of position and place of online content, created by the links among the Web Beings. Each Web Being is occupying a specific locus in the Web network. Identification in the Web Space is given by the URI namespace. Location is specified by a triplet of URIs, namely the URIs of the Web Being and the incoming and outgoing links. These links provide orientation by acting as a three-dimensional "geographic coordinate system" in the Web. The act of creation or deletion of a Web Being or a link, alters the Web Space. Hence, the evolving Web Space is fully describable by the lifetime processes of Web Beings and links. Except for the "book-keeping" clock time defined in Physics, time could be a series of choices in space. Web time is a series of choices (visits) in the Web Space that can be defined as Bergsonian durations, since visiting selections attach meaning and define casual relationships among Web Beings. This approach of time as duration is characterized by indeterminism, heterogeneity and irreversibility. In the Web, durations are becoming discoverable, observable, traceable, able to process and massive.

### 3.3 The "manna from heaven"

The study of codependence among Code and Users is really complicated. Initially, we suggest that on the first level the Web can be studied as the only existing system in the world. Human beings are communicating and working solely through and with the Web. A compassionate 'God' provides the necessary quantity of 'manna', fulfilling all human needs, with no cost and effort. This strong and unrealistic assumption will help us to comprehend bottom to top the moral values and their inter-connections to the complex actualities of the engineering principles. The analytical outcome of the first level will prepare us to study the effects of the Web in the entire human society. A characteristic domain of application of this assumption can be Net Neutrality issues. It will include the comparative analysis between established and emerging of new theories in the social, technological and economic domain. Analyzing the internal Web ethics at the first level will provide us with the necessary insights about neutrality as the interplay of core values and the engineering of the Web.

### 3.4 Technology, context and economy

The Web can be analyzed on three levels: the technological, the contextual and the economic, since they reflect its historical evolution from plain software to living ecosystem. The Web technology is built on the Internet, resulting huge amounts of data created by billions of Users (technology level). On top of this software, various new contexts have expanded initial functionalities. Context, being a set of tasks or a general framework of attitudes, enables Users to extent the range of information exchange and collaborative action, mainly through trust mechanisms (context level). The establishment of beliefs and attitudes regarding the trustworthiness of Users and associated Web Beings enabled the emergence of business models that are based on exchanges – financial or other – among Users (economic level).

Note here a point made by [30] who argues the importance of the distinction between trustworthiness/trust and reliability/reliance. He locates the distinction in the nature of the interactions between trustor and trustee. Where the interactions are 'static', we merely have a case of reliance (as someone may rely on a bridge that has been well-built, or on a clock that is correct). The emergence of trust out of reliance is an important signal for the move up from the technology level.

For Pettit, trust only comes when the interaction is interactively *dynamic* – i.e. trustworthy agents understand they are trusted, and trust gives them additional motive for behaving in a trustworthy manner. He argues on this basis that trust over the Internet (and *ipso facto* the Web) is impossible without supporting offline relationships and information, and therefore impossible on the 'manna from heaven' assumption discussed above. The reason for Pettit's rejection of trust as a possibility in this context is the fluidity of identity online – how could a trustor come to believe that a virtual contact fulfilled the requirements for interactive dynamism?

Without getting too deeply into this issue, [25] moves the focus for trustworthiness away from the trustee's attitude to the trustor, and toward the claims about her intentions, capacities and motivations the trustee makes. In particular, it is an attractive suggestion that the shift from reliability to trustworthiness happens as these claims become less deterministic, more implicit and less precise. There is no exact borderline or tipping point, but this conveys the importance of the agency and the choice for the trustee.

### 3.5 Hayek's theory of freedom

According to [17], "liberty" or "freedom" is defined to be the absence of coercion of some humans by other humans. This does not mean that one has unlimited options including all physical potentialities of the world. Likewise, it does not account for the internal states of being and any metaphysical notion of freedom or power. The main focus is on the mitigation of coercion as a set of restraints or constraints to human will, imposed by others. As Hayek explains (p.133), *"Coercion occurs when one man's actions are made to serve another man's will, not for his own but for the other's purpose. It is not that the coerced does not choose at all; if that were the case, we should not speak of his "acting."* Similarly, Hayek defines important facets of coercion like deception and fraud, as forms of controlling the information upon which a human counts; this information makes a human do what the deceiver wants him to do. Despite the fact that coercion suggests both the threat of inflicting harms and the intention thereby to cause certain outcomes, it does not necessarily involves all influences that humans can exercise upon the acting of others and acquire full control of the environment. Coercion is undesirable because it *"prevents a person from using his mental powers to the full and consequently from making the greatest contribution that he is capable of to the community."* (p.134). On the contrary, freedom is desirable *"because every individual knows so little and, in particular, because we rarely know which of us knows best that we trust the independent and competitive efforts of many to induce the emergence of what we shall want when we see it."* (p.29). It is freedom that releases the unforeseeable and unpredictable; these little accidents in human behavior, which are so vital for innovation. As Hayek argues (p.31) *"It is because we do not know how individuals will use*

*their freedom that it is so important."* and *"Freedom granted only when it is known beforehand that its effects will be beneficial is not freedom."* These accidents are the resultant of knowledge and attitudes, skills and habits, formed by human interaction and, in most cases, they do not simply occur but evolve. In order to flourish they must be supported by the existence of complementary concepts like some personal sphere, property, state, rules, competition and responsibility. The emergence of personal sphere and property assists individuals to avoid coercion from others. The only means to prevent coercion is the potential threat tied to coercion. States typically monopolize coercive power. In free societies, the State exercises minimal enforcement of coercive power, which nurtures individual creativity and competitive markets based on just distribution of property and responsible individual behavior. Particularly, *"Since coercion is the control of the essential data of an individual's action by another, it can be prevented only by enabling the individual to secure for himself some private sphere where he is protected against such interference. ... It is here that coercion of one individual by another can be prevented only by the threat of coercion assured free sphere."* (p.139). The acquisition of property is the first step towards the limitation of personal sphere and against coercive action. The next steps include the initiation of general rules governing the conditions under which behaviors and attitudes become part of such individual spheres (it is clear that carefully-crafted data protection rules are vital for both steps, which makes the lack of cooperation, or even of an agreed framework, between the EU and the US, not to mention India and China, all the more disturbing). It is crucial to ensure that the range and content of these rules is not determined by the deliberate assignment of particular things to particular persons. *"The decisive condition for mutually advantageous collaboration between people, based on voluntary consent rather than coercion, is that there be many people who can serve one's needs, so that nobody has to be dependent on specific persons for the essential conditions of life or the possibility of development in some direction. It is competition made possible by the dispersion of property that deprives the individual owners of particular things of all coercive powers."* (p.141). The degree of freedom in a society is directly related to the minimal enforcement of coercive power by the state according to general and no discriminative rules and the safeguarding of competitive market conditions. Competition as the existence of an efficient number of alternative offers is fundamental in the case of providing life-critical services. Generally, *"whenever there is a danger of a monopolist's acquiring coercive power, the most expedient and effective method of preventing this is probably to require him to treat all customers alike, i.e., to insist that his prices be the same for all and to prohibit all discrimination on his part. This is the same principle by which we have learned to curb the coercive power of the state."* (p.136). Having argued about the strategic role of state in minimizing coercion does not connotes that individuals enjoy only the opportunity and the burden of choice; it also highlights that individuals must accept the consequences of their choices and the resulting approbation or censure for them. In a free society freedom and responsibility should be interlocked.

## 4. THE WEB AS A SPACE OF FREEDOM

For many philosophers, freedom is not just one of the values but constitutes the source and prescribes the conditions of most moral values [3]. Hence, a theory about freedom is necessary in order to explore the internal Web ethics. In the present article, Hayek's analysis about freedom is adapted because is adequately consonant to the main architectural principles of the Web artifact, namely: lack of central authority, openness, variety of choices, distributed empowerment of individuals and liberal underpinning. Hayek's approach is not the only theory of freedom that can be used to analyze Web ethics. Its clarity and generality enable us to build a starting point that will be extended and refined with other theories to capture the ethical aspects of the Web.

Freedom creates more options to solve problems collectively and to innovate, but some of these options may be used in ways that cause coercion *("freedom-coercion" tradeoff)*. Thus, the question enveloping each theory is how to construct a system that selects, with minimum social cost which positive options to sacrifice in order to minimize coercion (or the dual problem). Hayek's approach could be considered to offer one of the systematic answers in this question. In particular, his theory is briefly transcribed as follows:

- *State* posses the monopoly to enforce coercive power through *General Rules*.
- *Personal Sphere* and *Property* counterweight *state* power.
- *General Rules* are enforced equally and describe the borderlines between *state and Personal Sphere*.
- *Property* is a basic realization of *General Rules*.
- *Competition* is possible by the *dispersion of Property*.
- *Mutually advantageous collaboration* is based on *Competition* in service provision.
- An effective *anti-monopolistic policy* is to require from the monopolist (including the state) to *treat all customers alike*.
- *Individuals* should be *responsible* and *accountable* for their actions.

In the following Subsections we consider some "freedom-coercion" tradeoffs on three levels of abstraction (technology, context, economy) according to Hayek's conceptualizations, in order to gradually build a set of important issues about living with the Web.

### 4.1 The Technological Web

The Web is an engineered artifact, not some natural phenomenon. It has been created as an Internet application and its building blocks are crafted in software code. In this sense, technological underpinnings are vital for its existence.

#### 4.1.1 Internet infrastructure

Internet has been evolved from communication architecture for computers to generative system for innovative software, basically because it was built on simple principles that transfer the power of choice to equally trusted single Users. The absence of central gate keeping and the unprecedented decentralized power in action is coming with two major costs: (a) inefficient personal identity management and thus, lack of security and (b) not guaranteed quality of transmission.

The notion of Internet freedom is related to the free access and inter-connection of any compatible software code developed by Users over the Internet network. Coercive powers are mainly arising due to badware applications (e.g. computer-zombies), traffic censorship (e.g. "Snooping" - accessing information within Internet packets [4]) and inadequate quality of transmission. Personal sphere for Internet Users is described by their IP address whenever they are connected to Internet. IP addresses are traffic data that can only be processed for certain reasons (e.g. payments). Ordinarily, they are considered by Data Protection Authorities and courts to be personal data, despite the fact that courts in some countries (e.g. France) have reached conflicting decisions [20].

O'Hara has argued that the revolutionary aspect of the World Wide Web is that it is a decentralised information structure. This democratic decentralisation is a key factor in the added value that the Web provides, because it facilitates the serendipitous reuse of information in new and unanticipated contexts. However its basic principle, of free flow of information packets and a very simple set of rules and standards underpinning these complex structures, is being undermined by attempts to restrict information flow. As use of the Web has spread, illiberal regimes feel threatened, but thanks to the hands-off approach of the 1990s, there are no affirmative globally-recognised principles governing the flow of information online. Currently, China is still focusing on a censorship-based approach to information control, using methods in direct opposition to the Web's essential governing principle of decentralisation. The liberalism of the Web has two distinct levels: first, the free flow of information and unrestricted linking helps make the valuable network; secondly the engineering principles of the Web facilitate the efficient flow of information and enables the basic structure to attain balance. In this way, ethical principles (and a strong stand on a political dispute) influence directly even Web infrastructure [26].

### 4.1.2 The case of Net Neutrality (NN)

The definition of NN and its technical consequences as Internet traffic subject to no hindrances could be further elaborated by using Hayek's ideas. The "first-come first-served" model with no other restriction is extended to Quality of Service (QoS) discrimination as long as there are no special and exclusive contracts at work (limited discrimination and QoS). Hence, in the one hand, no one may have exclusivity to end points, but on the other hand, anyone can pay to have higher QoS in its end point. Alternatively, limited discrimination without QoS tiering can be applied. According to some lawmakers in the US, QoS discrimination is allowed, subject to no particular charge for higher-quality service [10]. The underlying technical challenge is to engineer solutions that ensure NN in combination with higher QoS. This can be achieved by designing Internet infrastructure that allows for implicit traffic differentiation and prioritization of a select traffic, but without any kind of User, network operator or ISP intervention. Such a proposal, which involves an implicit kind of datagram separation rather than an a-priori explicit flow prioritization, is called FAN (Flow-Aware Networking) [19, 34]. FAN may ensure neutrality along with the awareness of QoS [9]. This is because it does not aim to explicitly categorize data flows in distinct classes (e.g. premium, basic), but only to create an occurrence, upon which the implicit separation will be performed solely based on the current link status (e.g. dataflow congestion, traffic bottleneck etc.). Therefore, all datagrams are forwarded unconditionally in the pipeline, but they are also "equal", subject to be separated or even dropped when the network tolerance demands it. The main advantage of FAN-based architectures is that they differentiate the data flow, taking into account only the traffic characteristics of the currently transmitted information. Hence, apart from data discrimination, it is not possible to comprehensively discriminate certain applications, services and end-Users. Such NN-QoS symbiosis does not violate NN and data discrimination principles. It however demands a global implementation approach in infrastructure level, involving common standards in prediction and limitation mechanisms for controlling the quality of transmitted information in the pipeline. The limitation mechanisms may provide a sudden separation of flow, but the decision should be made upon specific network tolerance metrics rather than individual properties of specific flows, such as "who" sends/receives a specific "class" of information.

### 4.1.3 The Web software

The notion of freedom in the Web software is to freely navigate, create and update Web Beings and links. Its cornerstones are universality, openness and separation of layers in engineering, editing, searching and navigating. [4] argues that *"Keeping the web universal and keeping its standards open help people invent new services."* Coercive powers can be directly injected into the network by Internet infrastructure (e.g. NN). Badware-infected Web Beings [41], central control and censoring of Web traffic are main sources of internal coercion in the Web. The emergence of "walled gardens" in cabled TV and Social Networks [4, 41] are based on isolated or malformed (i.e. without exclusive or open URI) Web Beings that strengthen coercive potential through privacy threats and fragmentation. Furthermore, any effort to manipulate for own benefit the results of indexing and searching processes (e.g. spamdexing [24]) is a form of coercion because it distorts searcheability and navigation.

Navigation in the Web space results in traffic. Web traffic is recorded in the Web Being's log file. Actually, this is the first time that humanity has introduced a universal event log in such a stratified and heterogeneous system. The resulting log file is under *common ownership* by design. Both the Editor who administers and updates the particular Web Being and the Navigator, who visits it, share the same information about this event. Although, the Editor has direct access to the log file residing in the Web server, the Navigator should install particular software to process the source file of his visiting history. Thus, this log file is the core architectural element that manifests the co-operative nature of the Web artifact and should be further analyzed. For the moment, legal and illegal cookies are censoring our moves with or without our consent [1] and "toolbar" applications exchange their services for recording all our navigation history.

During the first Web era, the majority of Users were Navigators and just a small portion of them was editing the Web. At the current Web 2.0 era, 70% of Users are both Navigators and Editors, who can easily edit, interconnect, aggregate and comment upon text, images and video. The underlying structure of the Web graph is characterized by four major characteristics: 1) on-line property (the number of Web Beings and links changes with time), 2) power law degree distribution with exponent higher than two, 3) small world property (the diameter is much smaller than the order of the graph) and 4) many dense bipartite sub-graphs [6]. In order for the Web to be an advantageous multi-purpose space, it should consist of a critical mass of Web Beings and links in an appropriate structure to facilitate navigation. Intuitively, it should be connected, not fragmented, to ease navigation from any Web Being to the entire network. The analysis of the interplay among functions, subsequent structures and moral values is an open question for internal Web ethics.

Treating all Navigators equally is an engineering principle. It is violated (or enriched) by profile customization. Treating all Editors alike is achieved through open technological standards developed by independent bodies (e.g. W3C). Public and private contribution to these institutions is necessary to sustain open and effective standards. Apart from the first class principles of universality, openness and separation, "quality-related" issues could be relevant to Web freedom if navigation and searching is severely degraded. Despite the fact that the explosion of bits in Web 2.0 increased the number of available Web Beings, incommoded the discovery of meaningful answers. This overload of unstructured content is partially tackled by Search Engines. Semantically structured data (aka Web 3.0) are engineered to anticipate it through machine-processable meaningful reasoning. The quality of content also includes factors like diversity, credibility, accuracy and informativeness of online content and stability of links.

## 4.2 The Contextualized Web

The Web became a techno-social space for innovation and inter-creativity because it has been transformed from a bulletin board to a context-aware system. It is not only the number of options the Web is providing, but also it is the quality and the usefulness of these options that matters. The Web context emerges as a bridge in the traditional public-private dichotomy. The *privatized (or publicized) space* arises between the private realm of intimacy and individualism and the public realm of citizenship and active participation for the societal good [29]. On the contrary, in the industrial economy, where consumers are mainly exercising the right to use resources, Web Users exercise the full range of property rights, namely: (1) to use, (2) to form, modify and substantiate (3) to benefit from use and (4) to transfer Web Beings.

Context, as a set of tasks or general framework of roles and attitudes, enables Users to extent the range of information exchange and collaborative action, mainly through trust mechanisms. For instance, in Web 2.0, what Users create is not simply content (e.g. reviews) but context. This new contextual framework emerges through the aggregation and collaborative filtering of personal preferences in massive scale [39]. More importantly, it facilitates connected Users to search and navigate the complex Web more effectively, amplifying incentives for quality. Of course, there are many open issues to be solved such as the fashioning of more effective forms of online identities and trusting processes. According to [25], trust is an attitude toward the trustworthiness of an agent. In our Web-only hypothetical world ("manna from heaven" assumption), agents are the Users who control specific Web Beings. Representations, intentions, capacities, motivations and contexts are established and expressed exclusively by Web technologies. Hence, freedom in the contextualized Web is to establish specific contexts in order to form beliefs and attitudes that some Users and their underlying Web beings are trustworthy. Coercive powers can arise from un-trustworthy technologies and governments, social hacking, badware and malicious representations.

However, it is also important to take account of the bad forms that trust can take [2]. The links between coercion and trust are sometimes uncomfortably close. Note, for example, that when [16] describes his theory of encapsulated trust informed by rational-choice ideas in social science, he argues that *"I trust someone if I have reason to believe it will be in that person's interest to be trustworthy in the relevant way at the relevant time … [and if that person] counts my interests as partly his or her own interests just because they are my interests"* (p.19). What strikes the reader is how close this definition of *trust* is to Hayek's definition of *coercion* quoted earlier.

This brings in Baier's notion of antitrust [2], where trust is harmful to the society at large. In this case the focus is on areas where trust shades into coercion, but it is clear that there are other spheres of life where freedom undermines trust, or allows corrosive examples of trust to emerge – cybercrime is an obvious example, where trust among criminals is essential to prevent police infiltration, and where trust among Web users is exploited by criminals. Baier's expressibility test [2] (pp.123-124) asserts that a trust relation is morally acceptable provided that the trustee may express her motives truthfully; this is an important insight, but it must be vulnerable to Pettit's worry that such expression, in the world we are envisaging, could only be mediated by Web technologies.

Nevertheless, communication is central to establishing trust, as Habermas argued [15], and so the rich connectivity of the Web is bound into its function. Antitrust and coercion may well be prices we have to pay for widespread and beneficial trust (repeating Hayek's point that freedom may at all times produce bad outcomes). The point of a Web ethics is to try to ensure not that antitrust happens, but that it is outweighed by beneficial trust to as great a degree as possible consistent with Hayekian notions of freedom.

## 4.3 The Economic Web

Most needs are better fulfilled through collective effort. In practice, incentives, capabilities, preferences and realizations of effort are heterogeneous and difficult to be synchronized. A powerful metaphor to achieve synchronization is setting efforts and the products of them under a common valuation scheme, a uniform *numeraire*. This numeraire is money, supported by a set of institutions and practices (e.g. the market). It is far beyond the scope of this paper to analyze related economic theory. We limit ourselves to the reassurance that economizing a system is an important factor for its viability, usability and development. The issues posed in preceding layers could be viewed through the economic aspect (e.g. NN as two-sided pricing [11]). The question is how the above-mentioned freedoms can be efficiently engineered and disseminated across Users in particular techno-social contexts.

The Web has not emerged as a business project with hierarchical structures. It has been crafted as a creative and open space of volunteers, predominantly outside traditional market and pricing systems. In our point of view, markets would have never invested such amounts in labor costs to develop this gigantic system. But to be fair, market mechanisms provided the necessary motives and tools to initiate a high-risk idea like Web. Furthermore, the lack of direct compensation and the temporal disconnection between effort and rewards are the shared characteristics among Peer, Procurement and Patronage production models. In the Web, Peer production has been established as a basic form of production, extending David's taxonomy [7] with the fourth P [37] .

The explosion of Web Users occurred as a result of symbiosis between non-financial and financial incentives [37]. Accordingly, freedom in the economic Web pertains to the removal of any possible barrier to economize. Each User should be allowed to apply any business models. Apart from the preceding levels, coercive powers are coming from two economy-related sources: the concentration of power in a minority of Web Beings and Users and the inability of some Users to benefit from the Web economy. As the economic Web grows, state faces unprecedented and complex trade-offs between private interest and social welfare. Three of those are referred as the "Link economy", the "App economy" and the excessive market power in Search Engine market. Recently, the formation of links, a fundamental characteristic of the Web, became the center of business controversies. As traditional content creators (e.g. TV) are losing a large part of their revenue streams from User-Generated substitutes (e.g. micro-blogs), the need for the institution of regulation issues in free reference linking appears. On the other hand, it is argued that Search Engines create exploitable traffic for content creators and that all online content must be open, with permanent links, so that it may receive in-links, since links are a key to securing efficiency in creating and finding information. However, the economic implications of reference links on attention and revenue have not been analyzed yet, despite their influence over consumer's utility, competition and social welfare. [8] concluded that: *"link equilibria often do not form, even though their formation can lead to higher aggregate profits and better content. This, in the view of the authors constitutes a negative side-effect of the culture of "free" links that currently pervades the web…"*

Despite the fact that Web 2.0 multiplied the pool of Users and content, the direct use of Web technologies has become shallower. Contrastingly to early stages of Web's inception, modern Users are mainly using the Web through established services (e.g. Search Engine, Social Network) and not directly, for instance, by creating their homepage or concentrating and controlling personal data in a privately owned domain. [4] reasons that the tendency for some companies to develop native applications for specific devices (e.g. "app stores") instead of Web applications sterilizes and fragments the Web. [32] demonstrates that the already large levels of concentration in the Web search market are likely to continue. He argues that since the market mechanism cannot provide socially optimal quality levels, there is space for regulatory engagements which may involve the funding of basic R&D in Web search, or more drastic measures like the division of Search Engines into "software" and "service" parts. It seems that massive use is coming with the cost of *centralization* of both traffic and data control. The balance point between innovation coming from large Web companies and innovation from single Users or voluntary groups should be thoroughly examined. In our point of view, this fast evolving centralization is directly against the core values of the Web ecosystem and must be addressed in the direction of transferring back to individual User part of data control. This can be achieved through technologies and business practices that are transparently enabling the User to process and economize personal data. In this campaign, the primer difficulties arising from the fact that now the Web is partly governed by economic forces and traditional institutions, which are characterized by irrelevant or conflicting moral principles. Therefore, one of the fundamental issues for Web ethics is to put this debate to the foreground through the employment of concrete architectural and policy structures (for example, with reference to the conditions, formats and licenses under which Public Sector Information for reuse is made available to citizens).

## 5. RESULTS AND DISCUSSION

We believe that the Web engineering principles are ethically-relevant and they should be systematically analyzed as such, in order to realize their potential in promoting human values. Web ethics raises the question about what could be a better future with the Web and how we can engineer it. As an emerging field of applied ethics, it discerns the core values of Web's inception and their evolution process in diverse social contexts. Our main arguments are based on the codependence of code and values. The Web is seen as a new form of existence [36] and it is assumed that it is the only existing system. The proposed methodology gradually analyzes the Web's complex reality by enriching underlying technology with human behavior aspects. Our three-levels analysis (technology, context and economy) reflects the historical evolution of the Web from software to a social ecosystem. As the concept of freedom is a prerequisite of most of the moral values, we introduce our methodology on internal Web ethics by demonstrating Hayek's theory of freedom in the three-levels analysis of the Web. We Hayek's approach because it reflects nicely the codependence among the architectural engineering principles of the Web and moral values. This correspondence can be summarized as follows:

o centralization of traffic and data control, rights on visiting log file, custom User profiles and interplay among functions, structures and moral values are directly connected to the quality of freedom in the Web,

o issues about freedom in lower levels of the Web ecosystem (i.e. technology) have crucial impact on the subsequent levels of higher complexity (i.e. context, economy) and

o generic freedom-coercion trade-offs are useful in framing the feasibility space but incomplete in treating more specific cases in practice (e.g. NN).

As the Web grows, it becomes essential to balance the need for efficient efforts and the stimulus for more competitors in creating and economizing content and search provision. A basic prerequisite in this effort is to identify and engineer its core moral values in order to account for an extensive range of User functionalities and pervasiveness in social discourse. This ongoing work can be further inspired by philosophical theories and historic periods [27] (pp.207-209). Also, it will be placed and compared with regards to relevant research about the interplay between technology and society. Providing deeper insights in Web ethics requires the supplementary specification of the suggested model with sound theoretical foundations and more realistic assumptions. Therefore, the next steps should include the enrichment of contextualized Web with theories and technologies about identity, privacy and trust. The study of the ethics of the economic Web should be extended to the study of inequality and distribution theories and detailed business models. During the next phase of this research project, the "manna from heaven" assumption will be relaxed and the three-levels model will be augmented by a fourth level to capture Web's interaction with other real systems. At a latter stage the Web ethics should be able to address more pragmatic questions like: "Can the Web protect itself as a liberal society? How do we manage online identities ethically? How can I deal fairly with people if I don't know their expectations? If I don't even know they are people? " [28]. How the Web's function, structure and evolution are affected by ethics?

The Web is a unique piece of technology not only because of its breakthrough technological innovation, but mainly because it provides a new basis for expressing human creativity, and reveals "inactive" parts of human nature. Apart from understanding its morality, it is an inspiring challenge to transfuse the essence of our experience and the values of the Web to reassess concepts like freedom, choice, participation, inequality and development. We agree with [40] that *"It is not just information that must be free, but the knowledge of how to use it. The test of a free society is not the liberty to consume information, nor to produce it, nor even to implement its potential in private world of one's choosing. The test of a free society is the liberty for the collective transformation of the world through abstractions freely chosen and freely actualised."* The role of Web ethics could be to elaborate and specify the motives and engineering of this new version of utopia.